\documentstyle[12pt]{article}
\def\BM#1{\mbox{\boldmath{$#1$}}}
\def\beqra{\begin{eqnarray}}
\def\eeqra{\end{eqnarray}}
\def\beqast{\begin{eqnarray*}}
\def\eeqast{\end{eqnarray*}}
\def\be{\begin{enumerate}}
\def\ee{\end{enumerate}}

\def\ul{\underline}
\def\sppt{Research supported in part by the
Robert A. Welch Foundation and NSF Grant PHY 9009850}

\def\utgp{Theory Group\\ Department of Physics \\ University of Texas
\\ Austin, Texas 78712}

\def\fnote#1#2{\begingroup\def\thefootnote{#1}\footnote{#2}\addtocounter
{footnote}{-1}\endgroup}

\def\beq{\begin{equation}}	\def\eeq{\end{equation}}
\def\haf{\frac{1}{2}}
\def\pa{\partial}

\def\rf#1{$^{#1}$ }

\begin{document}

\hfill{UTTG-11-92}

\vspace{24pt}

\begin{center}
{\bf   Three-Body Interactions Among Nucleons and Pions}
\vspace{36pt}

 Steven Weinberg\fnote{*}{\sppt.}

\vspace{24pt}
\utgp

\vspace{30pt}
{\bf Abstract}

\end{center}

\begin{minipage}{4.75in}
A chiral invariant effective Lagrangian may be used to
calculate the three-body interactions among low-energy pions
and nucleons in terms of known parameters.  This method is
illustrated by the calculation of the pion-nucleus
scattering length.

\end{minipage}

\vfill

\baselineskip=24pt
\pagebreak
\setcounter{page}{1}

Recent articles\rf{1,2} have described a systematic
effective Lagrangian framework for the calculation of
reactions involving arbitrary numbers of nucleons as well as pions of low
3-momentum.  To leading order in small momenta, the `potential' for such
reactions is given entirely by the tree graphs in which only two of the pions
and/or nucleons interact;  further, their interaction is calculated using the
original effective chiral Lagrangian\rf{3}, which consists of terms with only
the minimum numbers of derivatives or pion mass factors, supplemented by
contact
interaction terms among nucleons.   The corrections to these two-body
interactions of  second order in small momenta involve not only one-loop
graphs,
but also a large number of new terms\rf{4} in the Lagrangian with additional
derivatives, so many that not much can be learned about pion-nucleon or
nucleon-nucleon interactions in this way.  Fortunately, these two-body
interactions can instead be taken from phenomenological models that incorporate
experimental information on nucleon-nucleon, pion-nucleon, and pion-pion
scattering.  The only remaining contributions to the potential of the same
order
in small momenta consist of graphs in which {\em three} particles (or two pairs
of particles) interact, their interactions given by tree graphs calculated from
the original effective chiral Lagrangian.  Thus we can use the three-body
interactions calculated in terms of known parameters from the original
effective
chiral Lagrangians together with experimental data on two-body scattering to
calculate all corrections to the potential of first and second order in small
momenta.

This method will be illustrated here in the calculation of the amplitudes for
pion scattering on complex nuclei.  But first, a reminder of some generalities.

Consider the
amplitude for a process with $N_n$ nucleons and $N_\pi$
pions in the initial state and the same numbers of nucleons
and pions in the
final state, all with 3-momenta no larger than of order
$m_\pi$.    We wish to develop a perturbation theory for
this amplitude, based on an expansion in powers of the ratio
of these small momenta (and the pion mass) to some momentum
scale that is characteristic of quantum chromodynamics,
such as $m_\rho$.  In counting the number of powers of small
momenta in any given ``old fashioned'' (time-ordered)
diagram for this process, we must distinguish between energy
denominators of two types.  Those of the first type arise
from intermediate states that differ from the initial and
final states  in the number of pions and/or in the pion
energies, and are therefore of the order of the small
momenta or the pion mass.  The energy denominators of the
second type arise from intermediate states that differ
from the initial and final states only in the nucleon
momenta, and are therefore much smaller, of the order of the
nucleon kinetic energies.  A given graph is called  {\em
irreducible} if it contains only energy denominators of the
first type.  These are graphs for which the initial particle
lines cannot all be disconnected from the final particle
lines by cutting through any intermediate state containing
$N_n$ nucleons and either  all the initial or all the final
pions.    We shall consider disconnected as well as
connected irreducible graphs, because a general connected
graph is built up from a sequence of both  disconnected and
connected irreducible graphs interleaved with small energy denominators of the
second type.   (However in all graphs considered here, each of the initial
particle lines must be connected to one or more of the final
particle lines, and vice versa.) This sum of disconnected and
connected irreducible graphs is what was referred to above as the potential.

Because irreducible graphs do not contain  anomalously small
energy denominators of the order of nucleon kinetic
energies, it is easy to count the number $\nu$ of powers of
small momenta or pion masses in these graphs.  For an
irreducible graph with $V_i$ vertices of type $i$, $L$
loops, and $C$ separate connected pieces, the number of
powers of small momenta or pion masses is\rf{1,2}
\begin{equation}
\nu = 4-N_n-2C + 2L
+ \sum_i V_i\,\Delta_i\;,
\end{equation}
where $\Delta_i$ is an index for an interaction of type $i$,
given in terms of
 the number $n_i$ of nucleon field factors and the number
$d_i$ of derivatives (or powers of pion mass) in the
interaction, by
\beq
\Delta_i =d_i+\frac{1}{2}n_i-2\;.
\eeq
(In deriving this result, we count $-3$ powers of small
momenta for each line passing without interaction through
the diagram, because the associated momentum-space delta
function reduces the number of momentum factors in the total
connected amplitude by that amount.)

Eq. (1) is useful because chiral invariance rules out any
terms in the Lagrangian with $\Delta_i<0$.  It follows that for any
given number of external lines, the leading irreducible
graphs (those with smallest $\nu$) are the tree graphs (i.
e., $L=0$) with the maximum number $C$ of
connected parts, constructed solely from vertices with
$\Delta_i=0$.
The contribution of these vertices can be read off from the effective
interaction Hamiltonian (in the interaction picture):
\begin{eqnarray}
 H_{{\rm int},\Delta=0}& =& \haf (D^2 - 1) {\dot{\BM \pi}}^2
+ \haf (D^{-2}
- 1) \vec{\nabla}{\BM \pi}\cdot \vec{\nabla}{\BM \pi}
+\haf m_{\pi}^2 (D^{-1} - 1) {\BM \pi}^2 \nonumber  \\& &
+2F_{\pi}^{-4} (\overline{N}({\BM t} \times {\BM \pi})N)^2
\nonumber  \\& &
+\overline{N} \left[2F_{\pi}^{-1}g_A D^{-1}{\BM t}\cdot
(\vec{\sigma}\cdot \vec{\nabla}{\BM \pi}) +2F_{\pi}^{-2}D
{\BM t} \cdot ({\BM \pi} \times \dot{\BM \pi})\right]N
\nonumber  \\& & + \haf C_S(\overline{N}N)(\overline{N}
N) + \haf
C_T(\overline{N}\vec{\sigma}N)(\overline{N}\vec{\sigma}
N)\;.
\end{eqnarray}
which is derived from the most general chiral-invariant
Lagrangian with $\Delta_i = 0$
:
\begin{eqnarray}
{\cal L}_{\Delta=0}&=&- \haf D^{-2}\partial_{\mu}{\BM \pi}
\cdot \partial^{\mu}{\BM \pi}
-\haf D^{-1}m_{\pi}^2 {\BM \pi}^2 \nonumber  \\& & +
\overline{N}\left[i\pa_0 - 2D^{-1} F_{\pi}^{-2}{\BM t} \cdot
({\BM \pi} \times \partial_0{\BM \pi})
-m_N - 2D^{-1} F_{\pi}^{-1}g_A{\BM t} \cdot
(\vec{\sigma}\cdot\vec{\nabla}){ \BM \pi} \right]N \nonumber
\\& &
-\haf C_S (\overline{N}  N)(\overline{N}
 N) -\haf C_T(\overline{N}\vec{\sigma}  N)\cdot(\overline{N}
\vec{\sigma}  N)
\end{eqnarray}
where $g_A \simeq 1.25 $ and $F_{\pi} \simeq $ 190 MeV are
the
usual axial coupling constant and pion decay amplitude;
${\BM t}$ is the nucleon isospin matrix;
$C_S$ and $C_T$ are constants whose values can be inferred
from the singlet and triplet neutron-proton scattering
lengths; and $D \equiv 1+{\BM \pi}^2/F_{\pi}^2$.
(As discussed in ref. 2, terms involving time-derivatives of
the nucleon field are eliminated by a suitable redefinition
of that field, while corrections to the non-relativistic
treatment of the nucleon in (3) and (4) appear as terms in
the effective Hamiltonian and Lagrangian with $\Delta_i>0$.)
The number $C$ of connected parts is given its maximum value
$C=N_n+N_\pi-1$ by including only graphs (with one qualification to be
discussed later) for a single $\pi N$, $N N$, or $\pi \pi$ scattering, with
 all other lines passing without interaction through the diagram.

The corrections to these leading terms with only one extra
factor of small momenta (or $m_\pi$) arise from (a) tree
graphs, with the maximum number $C=N_n+N_\pi-1$ of connected
parts, that involve a single vertex (such as those
arising from non-zero $u$ and $d$ quark masses) with
$\Delta_i =1$, plus any number of vertices with
$\Delta_i=0$.  The next corrections, with two extra factors
of small momenta (or $m_\pi$), arise from (b) one-loop
graphs
with $C=N_n+N_\pi-1$ involving only vertices with
$\Delta_i=0$; (c) tree graphs with $C=N_n+N_\pi-1$ involving
either two vertices with $\Delta_i =1$ or one
vertex with $\Delta_i =2$ (which serve as counterterms
for the infinities encountered in one-loop graphs), as well
as any number of vertices with $\Delta_i=0$ ;
(d) tree graphs, constructed entirely from vertices with
$\Delta_i=0$, that have one less than the maximum number of
connected parts, i. e., with $C=N_n+N_\pi-2$.

As already mentioned, the vertices with $\Delta_i=2$ that contribute to
corrections of type (c) contain so many free parameters\rf4
that little of value can be learned by using the effective
Lagrangian to calculate these corrections.  On the other
hand, these corrections as well as the leading terms and the
corrections of types (a) and (b) all only contribute to the
maximally disconnected irreducible graphs, that consist of a connected piece
involving just two of the incoming nucleons and/or pions, plus disconnected
lines passing without interaction through the diagram for
all of the other incoming nucleons and pions.  But instead
of trying to use the effective Lagrangian to calculate such
two-body interactions, we can draw on various
phenomenological models that incorporate not only chiral
symmetry but the whole body of present experimental
information about low energy nucleon-nucleon,
nucleon-pion, and pion-pion scattering.

There remain only the corrections of type (d), with
$C=N_n+N_\pi-2$.  These are to be calculated from tree
graphs involving only the $\Delta_i=0$ Hamiltonian (3),
which involves no unknown parameters.  These corrections arise
from graphs that either consist of (d1) a connected
piece involving just {\em three} of the incoming nucleons
and/or pions, or (d2) two connected pieces each involving
just two of the incoming nucleons and/or pions, plus in both
cases disconnected lines passing without interaction through
the diagram for all of the other incoming nucleons and
pions.  Graphs of type (d2) may, like the graphs of types (a), (b), and (c),
be taken from suitable phenomenological models based on experimental
information  about two-body scattering processes.  This leaves only the
three-body graphs of type (d1),  which can be calculated from first principles
in terms of known constants.

Let's first see how this applies to processes involving only
nucleons.  Multinucleon scattering amplitudes and
bound-state wave functions are found by solving an inhomogeneous
Lippman-Schwinger or homogeneous Schr\"{o}dinger equation
with the effective potential taken as the sum  of
irreducible graphs.  The graphs for the three-nucleon terms
in the effective potential are shown in Figures (1) and (2).
A cancellation (to leading order in small momenta) was noted
in
reference 2  among the graphs of Figure (1), the only graphs
that involve the part of the effective Hamiltonian (3) that
is non-linear in the pion field.  It is instructive to look
at the reason for this cancellation.  These graphs all
involve a single quadratic interaction
$2F_{\pi}^{-1}g_A \overline{N}\, {\BM t} \cdot ({\BM \pi}
\times \dot{\BM \pi})N$
plus linear interactions of the two pion fields in this
interaction with the other two nucleons.  In each individual
time-ordered graph, the time derivative in the quadratic
interaction makes a contribution of the order of a pion
energy.  However, by summing up the old-fashioned graphs for
all the time-orderings of these three vertices,
we obtain a {\em Feynman diagram} in which energy is
conserved at each vertex, so that the time-derivative yields
a difference of  nucleon kinetic energies, smaller by a
factor at most of order $m_\pi/m_N$.

This leaves the 3-nucleon graphs of Figure (2).  These are
genuine contributions to what we have defined as the
3-nucleon potential, but they involve only the contact and
pion-exchange nucleon-nucleon interactions, and their effect
is actually cancelled by terms in the expansion of the
reducible three-nucleon graphs of
Figure (3) in powers of the ratio of nucleon to pion kinetic
energies.
Again, the reason for this cancellation is not hard to find.
Although in Figure (2) we are not summing over all time
orderings, so that these graphs do not make up a complete
Feynman diagram, the sum of all the time-ordered graphs of
Figures (2) {\em and} (3)  makes up several complete Feynman
diagrams, in which energy denominators are replaced with
pion and nucleon propagators, and energy is conserved at
each vertex.  Since the virtual pion energies
in these Feynman diagrams are equal to differences of
nucleon kinetic energies, and hence negligible compared with
the virtual pion 3-momenta $\vec{q}$, the pion propagators
in these diagrams are just $(\vec{q}{}^2 + m_\pi^2)^{-1}$.
But these Feynman diagrams with such pion propagators are
just what we would get from the old-fashioned diagrams of
Figure (3) if we were to neglect nucleon kinetic energies in
energy denominators for states containing a pion.
Thus we may calculate the multi-nucleon potential to second
order in small momenta by ignoring nucleon kinetic energies
in the energy denominators of the
leading pion-exchange contributions to the potential, {\em
and} ignoring the three (or more) - nucleon contributions
altogether.  This is more or less what nuclear physicists
have always done anyway.\fnote{*}{I am grateful to J. Friar
for pointing out that in some treatments of the nuclear
three-body problem the pion exchange forces are calculated
neglecting nucleon kinetic energies in energy denominators,
and that the corrections to this approximation are of the
same order as the other corrections considered in this
work.}

The three-body forces are  more interesting in processes
involving a pion.  For definiteness, consider the low-energy
elastic scattering of a pion from a nucleus of nucleon
number $A$.  General considerations of scattering theory
tell us that the S-matrix element for this process is simply
given by the matrix element between nuclear wave
functions of the sum of all irreducible graphs with $N_n=A$
and $N_\pi=1$.  (In applying the effective chiral Lagrangian
to such processes we are making use of the fact that typical
3-momenta of nucleons in nuclei are of order $m_\pi$ or
less.)  The leading irreducible graphs are those in which
the pion scatters off a single nucleon, evaluated using the
$\Delta_i=0$ vertices in the tree approximation.\fnote{**}{There are
also nominally
leading terms in which the
incoming pion is absorbed by one nucleon and
the outgoing pion is emitted by another, but when these are
summed over different time-orderings they cancel.  Again,
this is because summing over time-orderings yields a Feynman
diagram in which energy is conserved, but energy cannot be
conserved in the emission or absorption of a single real
pion by a single nucleon.}  To second order in
small momenta, the corrections to these leading terms arise
from corrections to the pion-nucleon scattering amplitude (from
loop graphs and
from vertices with
$\Delta_i=1,2$) which can be taken from
phenomonological models of pion-nucleon scattering, together
with connected three-body interactions among two nucleons
and the pion, calculated from tree graphs evaluated with the
$\Delta_i=0$ vertices in Eq. (3).
The graphs for these three-body interactions are shown in
Figure 4.

This is a lot to calculate, but the problem becomes much
simpler if we restrict our attention to the pion-nucleus
scattering length, for which the incoming and outgoing pion
have vanishing 3-momenta.  The leading terms as well as the
corrections to pion-nucleon scattering give a scattering
length that (apart from reduced-mass corrections) is just
the sum of the scattering
lengths on the individual nucleons.  This leaves only the
three-body irreducible graphs, of which the only ones that
survive in the limit of vanishing external pion 3-momenta
are those shown in Figures 4(a) to 4(f).

 It is easiest to calculate the contributions of Figures
4(a)-4(c) and 4(f) by noting that the sum over time
orderings in graphs of each type [and lumping together graph
4(f),   produced by the interaction term $2F_{\pi}^{-4}
(\overline{N}({\BM t} \times {\BM \pi})N)^2$ in the
Hamiltonian (3), with the other graphs]
must give the same result as the  complete Feynman diagrams
of type 4(a) - 4(c) calculated from the Lagrangian (4)
[which does not contain the interaction $2F_{\pi}^{-4}
(\overline{N}({\BM t} \times {\BM \pi})N)^2$.]  The other
graphs, 4(d) and 4(e), are not summed over all time-orderings
(because the sum  would include reducible as well as
irreducible graphs) and so their contributions must be
calculated using old-fashioned perturbation theory.   These
contributions to the pion-nucleon scattering length are:
\beqra
&& a^{[4(a)]}_{ab}= \frac{m_\pi^2}{2\pi^4 F_\pi^4
(1+m_\pi/m_d)}\sum_{r<s}\left<\frac{1}{\vec{q}_{rs}{}^2}
\left(2{\BM t}^{(r)}\cdot {\BM t}^{(s)}\delta_{ab}-
t_a^{(r)}t_b^{(s)}-
t_a^{(s)}t_b^{(r)}\right)\right>\;,\nonumber\\&&{}
\\&&
a^{[4(b)]}_{ab}= -\frac{g_A^2\delta_{ab}}{2\pi^4 F_\pi^4
(1+m_\pi/m_d)}\sum_{r<s}\left<{\BM t}^{(r)}\cdot {\BM
t}^{(s)}\frac{\vec{q}_{rs}\cdot\vec{\sigma}{}^{(r)}\;
\vec{q}_{rs}\cdot\vec{\sigma}{}^{(s)}}{\vec{q}_{rs}{}^2+
m_\pi^2}\right>\;,
\\&&
a^{[4(c)]}_{ab}= \frac{g_A^2}{2\pi^4 F_\pi^4
(1+m_\pi/m_d)}\nonumber\\&&\qquad\times\;\sum_{r<s}\left<
\frac{\left[{\vec{q}_{rs}{}^2{\BM t}^{(r)}
\cdot \BM
t}^{(s)}\delta_{ab}+m_\pi^2(t_a^{(r)}t_b^{(s)}+t_a^{(s)}t_b^
{(r)})\right] \vec{q}_{rs}\cdot\vec{\sigma}{}^{(r)}\;
\vec{q}_{rs}\cdot\vec{\sigma}{}^{(s)}}{(\vec{q}_{rs}{}^2+
m_\pi^2)^2}\right>\;,
\\&&
a^{[4(d,e)]}_{ab}=\frac{g_A^2 m_\pi}{8\pi^4 F_\pi^4
(1+m_\pi/m_d)}
\sum_{r<s}\left<({\BM t}^{(r)}+{\BM t}^{(s)}) \cdot ({\BM
t}^{(\pi)})_{ab}\frac{\vec{q}_{rs}\cdot\vec{\sigma}{}^{(r)}\;
\vec{q}_{rs}\cdot\vec{\sigma}{}^{(s)}}{(\vec{q}_{rs}{}^2+
m_\pi^2)^{3/2}}\right>\nonumber\\&&{}
\eeqra
where subscripts $a,b$ are pion isovector indices; $r,s$
label individual nucleons;   $\vec{q}_{rs}$ is the momentum
transferred between nucleons $r$ and $s$ in their
interaction with the pion; $\vec{\sigma}{}^{(r)}$ and
${\BM t}^{(r)}$ are the Pauli spin vector and isospin vector
of nucleon $r$; and $(t_c^{(\pi)})_{ab}=-i\epsilon_{abc}$ is
the pion isospin vector.
Note that as a result of a partial cancellation between (6) and (7),
the integrand in the sum of these averages vanishes for
$\vec{q}_{rs}\rightarrow \infty$, which makes the result
less sensitive to the behaviour of the nuclear wave function
at small internucleon separation.\fnote{\dagger}{This cancellation was
noted by Robilotta and Wilkin\rf{5} in the case of pion-deuteron scattering.
They used a different definition of the pion field, so their results for
diagrams 4(b) and 4(c)  were different from (6) and (7), but the sum of their
results agrees with what would be found for  this process from the sum of (6)
and (7).}  To second order in small momenta, the pion-nuclear scattering length
is
\beq
a_{ab}=\frac{1+m_\pi/m_N}{1+m_\pi/Am_N}\sum_r
a^{(r)}_{ab}+a^{[4(a)]}_{ab}+a^{[4(b)]}_{ab}+a^{[4(c)]}_{ab}
+a^{[4(d,e)]}_{ab}
\eeq
where $a^{(r)}_{ab}$ is the pion scattering length on the
$r$'th nucleon.

This all becomes much simpler in two special cases.  One is double
charge-exchange scattering,  $\pi^+ +N\rightarrow \pi^- + N'$, where the
scattering lengths $a^{[r]}_{ab}$ as well as the  corrections (6) and (8)
vanish.  The other, on which we shall concentrate here,  is pion scattering on
an isoscalar nucleus.  Here
$t_a^{(r)}t_b^{(s)}+t_a^{(s)}t_b^{(r)}$ may be replaced with
$\frac{2}{3}\delta_{ab}\, {\BM t}^{(r)}\cdot {\BM t}^{(s)}$,
and Eq. (8) vanishes.  More important, the contributions of the
nominally leading terms in the
pion-nucleon scattering lengths vanish, because they
involve an expectation value of $\sum_r {\BM
t}^{(r)}\cdot{\BM t}^{(\pi)}$, which vanishes in any
isoscalar nucleus.  The first term in (9) arises only from ``$\sigma$-term''
 corrections to
the pion-nucleon scattering lengths, and is therefore relatively small,
making it feasible to compare calculations of the corrections
considered here with experimental measurements of the
pion-nuclear scattering lengths.

This may be illustrated in the paradigmatic case of
pion-deuteron scattering.  To evaluate the two-body terms here we
need to use isotopic spin invariance to derive the
pion-neutron
scattering lengths from measured values of the $\pi^+p$ and
$\pi^-p$ scattering lengths.  This is not entirely
straightforward, because we are interested here in the
relatively small corrections to the leading soft-pion
results for which $a_{\pi p}+a_{\pi n}=0$, and these
corrections arise in part from ``sigma terms'' proportional
to $u$ and $d$ quark masses that do not even approximately
conserve isospin.  Fortunately to first order in quark
masses the isospin violation in the sigma terms affects only
processes involving at least one neutral pion,\rf6 so that
isospin relations {\em can} be used to calculate $a_{\pi
n}$.  This gives the two-body terms in the $\pi-d$
scattering length as\rf7
$\frac{1+m_\pi/m_N}{1+m_\pi/m_d}[a_{\pi p}+a_{\pi n}]=-
(0.021\pm 0.006)m_\pi^{-1}$.  Shifting to coordinate space,
the remaining corrections are given by:
\beq
 a^{[4a)]}=-\frac{ m_\pi^2}{\pi^2 F_\pi^4
(1+m_\pi/m_d)}\int^\infty_0 \frac{(u^2+w^2)}{r}\;dr\;,
\eeq
and
\beqra
 a^{[4(b,c)]}&=&\frac{ m_\pi^2 g_A^2}{3\pi^2 F_\pi^4
(1+m_\pi/m_d)}\,\left[\frac{1}{4}\int^\infty_0
(u^2+w^2)\left(\frac{1}{r}-\frac{m_\pi}{2}\right)e^{-m_\pi
r}\;dr\right. \nonumber\\&&\left.-\int^\infty_0
\left(\frac{uw}{\sqrt{2}}-
\frac{w^2}{4}\right)\left(\frac{1}{r}+m_\pi\right)\,e^{-
m_\pi r}\;dr\right]
\eeqra
where $u$ and $w$ are the s-wave and d-wave parts of the
deuteron wave function, normalized so that
\beq
\int^\infty_0 (u^2+w^2)\;dr=1
\eeq
The rescattering term (10) [but not (11)] has been previously considered in
the books of Eisenberg and Koltun and Ericson and Weise.\rf7
Because of the anomalously large radius of the deuteron, this term is
considerably larger than the remaining three-body term (11), so it should be
calculated including first-order corrections to the pion-nucleon
scattering vertices in Figure 4(a).  Fortunately these
corrections can be taken from the measured values of the
scattering lengths.\rf8  In this way one finds that\rf7
$a^{[4(a)]}=-(0.026\pm 0.001) m_\pi^{-1}$.
The remaining three-body terms (11) are
calculated\fnote{\ddagger}{The calculation of the integrals in
Eqs. (10) and (11) was carried out by R. C. Mastroleo and U.
van Kolck, using the Bonn wave function for the deuteron.}
to be $a^{[4(b,c)]}=-0.0005 m_\pi^{-1}$ (mostly arising from the
interference between s-wave and d-wave parts of the wave
function), in agreement with the numerical result quoted in reference 5.  This
is small compared with the uncertainties in other terms, and so may be
neglected here, though this may not be the case for pion scattering on heavier
nuclei.  This justifies the final theoretical result of reference 7,
 $a_{\pi d}=-(0.050\pm 0.006)m_\pi^{-1}$, which is in good agreement with the
experimental value $-(0.056\pm 0.009)m_\pi^{-1}$.
Although the use of chiral effective Lagrangians has turned out here only to
confirm previous calculations of pion-deuteron scattering as well as nuclear
binding,  the systematic counting of momentum factors in chiral perturbation
theory has proved its value in explaining (as previous calculations did not
explain) just  why it is correct to consider only certain graphs and certain
terms in the effective Lagrangian.

I am grateful for discussions with C. Dove, J. Friar, A.
Gleeson, C. Ordo\~{n}ez, U. van Kolck, and J. D. Walecka.
\pagebreak

\noindent
{\bf References}

\begin{enumerate}
%1
\item S. Weinberg, Physics Letters B 251 (1990) 288.
%2
\item S. Weinberg, Nuclear Physics B 363 (1991) 3.
%3
\item S. Weinberg, Phys. Rev. Lett. 18 (1967) 188 ; Phys. Rev. 166 (1968) 1568
{}.
%4
\item C. Ordo\~{n}ez and U. van Kolck, Texas preprint
UTTG-01-92, to be published in Physics Letters B291 (1992).
%5
\item M. R. Robilotta and C. Wilkin, J. Phys. G: Nucl.
Phys., 4 (1978) L115.  Also see H. McManus and D. O. Riska, Phys. Lett. 92B
(1990) 29.
%6
\item S. Weinberg, in {\em A Festschrift for I. I. Rabi},
Transactions of the N. Y. Academy of Sciences 38 (1977) 185.
%7
\item  J. M. Eisenberg and D. S. Koltun, \ul{Theory of Meson
Interactions with Nuclei} (Wiley-Interscience, New York,
1980); T. Ericson and W. Weise, \ul{Pions and Nuclei}
(Oxford University Press, Oxford, 1988).
%8
\item V. M. Kolybasov and A. E. Kudryavtsev, Zh. Eksper.
Teor. Fiz. (USSR), 63 (1972) 35; Sov. Phys. JETP, 36 (1973)
18.

\end{enumerate}
\end{document}